\documentclass[final,1p,times]{elsarticle}

\usepackage{amssymb}
\usepackage{indentfirst}
\usepackage{amsmath}
\usepackage{amssymb}
\usepackage{latexsym}
\usepackage{float}
\usepackage{graphics}
\usepackage{graphicx,float}
\usepackage{indentfirst}
\usepackage[colorlinks,hyperindex]{hyperref}

\journal{...}

\begin{document}

\begin{frontmatter}



\title{An Application of Lorentz Invariance Violation in Black Hole Thermodynamics}


\author[a]{Guo-Ping Li}
\author[a,b]{Jin Pu}
\author[b]{Qing-Quan Jiang\corref{mycorrespondingauthor}}
\cortext[mycorrespondingauthor]{Corresponding author}
\ead{qqjiangphys@yeah.net}
\author[a]{Xiao-Tao Zu}

\address[a]{School of Physical Electronics, University of Electronic
Science and Technology of china, Chengdu, Sichuan 610054, People's Republic of China}
\address[b]{College of Physics and Space Science, China
West Normal University, Nanchong, Sichuan 637002, People's Republic of China}

\begin{abstract}
In this paper, we have applied the Lorentz-invariance-violation (LIV) class of dispersion relations (DR) with the dimensionless parameter $n=2$  and the ``sign of LIV'' $\eta_+ = 1$, to phenomenologically study the effect of quantum gravity in the strong gravitational field. Specifically, we have studied the effect of the LIV-DR induced quantum gravity on the Schwarzschild black hole thermodynamics. The result shows that the effect of the LIV-DR induced quantum gravity speeds up the black hole evaporation, and its corresponding black hole entropy undergoes a leading logarithmic correction to the ``reduced Bekenstein-Hawking entropy'', and the ill defined situations (i.e. the singularity problem and the critical problem) are naturally bypassed when the LIV-DR effect is present. Also, to put our results in a proper perspective, we have compared with the earlier findings by another quantum gravity candidate, i.e. the generalized uncertainty principle (GUP). Finally, we conclude from the inert remnants at the final stage of the black hole evaporation that, the GUP as a candidate for describing quantum gravity can always do as well as the LIV-DR by adjusting the model-dependent parameters, but in the same model-dependent parameters the LIV-DR acts as a more suitable candidate.

\end{abstract}


\end{frontmatter}


\section{Introduction}
\label{sec:intro}
\setlength{\parindent}{2em}
In the Special Relativity, the speed of light $c$ in vacuum is treated as a constant for all inertially moving observers \cite{Jacobson}. The combinations of $c \left( {2.998 \times {{10}^8}m \cdot {s^{ - 1}}} \right)$, the plank constant $h\left( {6.626 \times {{10}^{ - 34}}J \cdot s} \right)$ and the gravitational constant $G\left( {6.673 \times {{10}^{ - 11}}N \cdot {m^2} \cdot k{g^{ - 2}}} \right)$ give rise to a length scale  $L_p= \sqrt{{G h}/{(2 \pi c^3)}}\approx 1.62\times10^{-35} m$ or its inverse ${E_p} \approx 1.22 \times {10^{28}}eV$, which is now named as the ``Planck length'' \cite{Giovanni1}. So far, this scale has been always considered as a boundary of the classical and quantum description of modern physics.
On the other hand, some attempts to combine gravity and quantum mechanics have suggested that the ``spacetime-foam'' picture should not be of a classical smooth geometry, but it should be quantized at a very short distance scale \cite{Townsend,Giovanni5}. This scale is always expected to be the so-called Planck length ${L_p}$ by the reason that both the Plank constant $h$ and the gravitational constant $G$ are the essential components of ${L_p}$ \cite{Giovanni0}. Therefore, the Planck length ${L_p}$ presents the strength of quantum gravity, even though it is extremely difficult to test in experiments \cite{Giovanni1}. If this worked, it suggests that the Planck-scale physics effect must naturally arise in various theories of quantum gravity.
However, in the Lorentz gauge theory-the Special Relativity (SR), the invariant Planck length ${L_p}$ is described in conflict with the effect of the Lorentz contraction. To avoid this discrepancy, Amelino-Camelia has introduced the Planck length ${L_p}$ as an observer-independent constant into the SR \cite{Giovanni2}. Now, this deformed SR, including two observer-independent constants, is always called as the Double Special Relativity (DSR) \cite{Giovanni2}. In the DSR, the modified dispersion relation that ensures the two observer-independent constants, i.e. $L_p$ and $c$, would lead to the Planck-scale departure from the Lorentz symmetry, which is also referred to as the Lorentz invariance violation of dispersion relations (LIV-DR). Our above discussions on the LIV are mainly based on the hypothesis of spacetime, but however, there also are some other popular ways to explore the LIV in the theoretical physics, such as the loop gravity \cite{Alfaro}, the foamy structure of spacetime \cite{Garay}, the torsion in general relativity \cite{Yan} and the vacuum condensate of antisymmetric tensor fields in string theory \cite{Kostelecky}. Using these methods, much fruit has been achieved \cite{Alfaro,Garay,Yan,Kostelecky}. Meanwhile, several other approaches to quantum gravity also suggest that, there maybe a Lorentz violation microscopic structure of spacetime, for instance the discreteness \cite{Gambini}, the non-commutativity \cite{Carroll} and the extra dimensions \cite{Burgess}. In a word, all of these theoretical observations tell us the fact that the LIV might be acted as a relic probe for quantum gravity.

Turning our attentions to the LIV on experiments, it is well-known for us that, the suddenly intense flashes of Gamma-ray bursts (GRBs) that originates from the distant galaxies with cosmological distances can give rise to the photons, which up to $\sim Tev$ energy \cite{Giovanni3}. When the Hamiltonian equation of motion $\dot{x_i} = {\partial H}/{\partial p_i}$ still holds true at least approximately, then the speed of those photons could have an energy dependence from the expression $v = {\partial E}/{\partial p}$ as a result of LIV \cite{Giovanni6}. That is to say, these simultaneously emitted photons should have a small energy-dependent time-of-arrival differences ($\Delta t$, namely the so-called ``time delay'') when they travel over very long distances to reach our observatories. At present, it is generally believed that GRBs can be regarded as an ideal object to observe the possible minuscule effects of LIV \cite{Giovanni3,Giovanni6}.
From the Fermi observations, various constraints on LIV parameters from the recent observation(such as, GRB090510 \cite{GRB2,GRB3}, GRB080916C \cite{GRB4}, GRB090902B \cite{GRB5} and GRB090926A \cite{GRB6}) has been worked out in \cite{GRB3,GRB7}.
By further considering the intrinsic time lag, effects of LIV from 8 data of bright GRBs with inclusion of GRB100414A and GRB130427A$^a$ has been carefully addressed by Zhang \cite{GRB1}.
Furthermore, because gamma-ray bursters can also emit a large number of high-energy neutrinos, which have the energy range between $10^{14} \sim 10^{19} eV$, Jacob suggested that one can employ neutrinos to replace photons in analysis of time-of-arrival difference that produced by LIV \cite{Jacob}. If this worked, it would enable us to generically set constraints on LIV at levels, which other methods can not be reached. Meanwhile, some other similar and remarkable phenomenons in experiment have been observed, in which constraints on the LIV attracted a lot of attentions \cite{Other1,Other2,Other3,Other4,Other5,Other6,Other7,Other8,Other9,Other10,Giovanni4}. In a word, the Lorentz invariance violation typically at the Planck scale has been received more and more interests both theoretically and experimentally in recent years.

For simplicity, we here consider the preferred frames in which dispersion relation break boost invariance but preserve rotation invariance. In this case, by considering the phenomenological levels, the Lorentz invariance violation (LIV) class of dispersion relations that maybe induced by several approaches of quantum gravity, can be expected to take the following form for the massive particles \cite{Giovanni3},
\begin{equation}\label{n}
{E^2} = {p^2} + {m^2} - {\eta _ \pm }{p^2}{\left( {\frac{E}{{{\xi _n}{M_{QG}}}}} \right)^n}.
\end{equation}
We should note that this relation can be considered only when it occurs at high energy scales. Also, it is only convenient for us to choose the form (\ref{n}) to work in the current context. In the Standard Model Extension (SME), which parametrizes the most general effective field theory with Lorentz violating operators, other different forms of LIV-DR can and do arise \cite{add1,add2}. The quantity $E,m$ and $p$ denote the energy, the mass and the 3-component momentum of particles, respectively. ${\eta _ \pm }$ is the ``sign of the LIV'' \cite{Giovanni3}, and the sign $\eta_+=1$ represents a subluminal correction and the sign $\eta_-=-1$ denotes a superluminal one \cite{Jacob}. However, it's worth noting that one should employ the sign $\eta_+=1$ to coincide with the experimentally astrophysical phenomena, which is emphasized by Jacob and Amelino-Camelia \cite{Giovanni3,Jacob}. The energy scale of quantum gravity ${E_{QG}}$ can be expected to be in the neighborhood of the Planck energy scale $({E_p} \approx {10^{28}}eV)$, which results into a nonvanishing contribution to the usual dispersion relation \cite{Biesiada}. Later on, some constraints on the quantum-gravity energy scale have been worked to show ${{M_{QG}} > 0.3{M_p}} $ \cite{Stecker1}. The coefficient ${\xi _n}$ is a dimensionless parameter, and lower bounds from gamma-ray bursts (GRBs) has been obtained for $n = 1$ case ${\xi _1} \ge 0.01$, and ${\xi _2} \ge {10^{ - 9}}$ from flaring active galactic nuclei (AGNs) for $n = 2$ case \cite{Jacob,Biller}. Here, $n$ is also a dimensionless parameter, usually appearing as an integer, and any noncommutative geometry can give a definite value of itself. Also, different values of $n$ are suppressed by ${M_p}$, denoting different magnitudes of corrections \cite{Giovanni3}. In our consideration, values at $n \le 0$ should be excluded in high energy scale, because huge deviations at low energies would be happened at this levels and led a strong limit \cite{Guo}. In the case of $n \ge 3$, the quantum-gravity corrections would be too small to be observed \cite{Giovanni3,Giovanni4}. As a result, more and more researches have confidence in the values of $n = 1$ for linear suppression and $n=2$ for higher-price quadratic suppression by the Planck mass ($1/{M_p}$ or ${L_p}$). In the framework of space-time foam Liouville-string model, the deformed dispersion relation with parameter $n=1$ has been introduced, and another case $n=2$ would be expected to be found in loop quantum gravity \cite{Ellis1}. At the phenomenological levels, it is very important to fix the value of $n$ since the difference of the LIV effect between $n=1$ and $n=2$ becomes obvious at high energy scale. For $n=1$, Jacobson has shown that the observation of $100$-$MeV$ synchrotron radiation from the Crab Nebula provides a new constraint $\eta > -7 \times 10^{-8}$ on the electron parameter $\eta$ in the context of the effective field theory \cite{Jacobson}\footnote{The coefficient $\eta$ can be found in \cite{Jacobson}, which is not same as $\eta_{\pm}$ in equation (\ref{n}) of this paper.}. Clearly, this condition is so strong that Jacobson concludes that quantum gravity scenarios implying this sort of the Lorentz violation are not viable, and the fact that $n=1$ was ruled out might consequently be a result of the CPT symmetry\footnote{As described in Ref. \cite{Bernabeu}, the symmetry under the combinations of charge conjugation (C), parity(P), and time reversal (T) transformations, at present appears to be the only discrete symmetry of Quantum Mechanics respected in nature experimentally.}, rather than the Lorentz symmetry. Later, by reviewing Jacobson's paper \cite{Jacobson}, Ellis has reported that the synchrotron constraint provided by the Crab Nebula on the electron's dispersion relation implies that the situation $n \leq 1.74$ for the electron should be removed \cite{Ellis2}. In brief, we have good reason to believe that if one expect equation (\ref{n}) to act as a candidate of quantum gravity, we should choose the case $n=2$ with form of constraining the Lorentz violation suppressed by the second power, ${E^2}/({\xi_2^2 M_p^2})$. However, one may argue that it would be impossible to test this quadratically-suppressed LIV effects in experiments. To answer this question, Amelino-Camelia have observed that the photon and neutrino observatories do have the required capability for providing the first elements of an experimental programme. For $n=2$, in the spirit of the time-of-arrival analysis of gamma-ray bursts, especially when we compared the time of arrival of these neutrinos emitted by gamma-ray bursts with the corresponding time of arrival of low-energy photons, one find that the expected time-of-arrival difference can reach $\Delta t \sim 10^{-6}s$. This remarkable LIV effects in the case $n=2$ are within the realm of possibilities of future observatories, as described by Amelino-Camelia \cite{Giovanni3}. Based on the above viewpoints, the most popular LIV-DR should be always fixed to the case of $n=2$, which is usually expressed as
\begin{equation}
{E^2} = {p^2} + {m^2} - {p^2}E^2 l_p^2, \label{Jeq1}
\end{equation}
where $l_p={L_p}/{\xi_2}$ and $ {L_p} = {1}/{M_p}$ is the Planck length.

A black hole as a special object with the strong gravitational field, has many interesting thermodynamic properties such as a negative heat capacity and so on. So, when the particle is gradually emitted from the black hole horizon, the black hole temperature becomes higher and higher till the final stage of the black hole evaporation. Thus, the energy of the emitted particle is getting higher and higher during black hole emission. As a result, the quantum effect of gravity becomes more and more important during the study of black hole radiation. The introduction of gravity into quantum theory brings an observer independent minimum length scale. A minimal length also occurs in string theory \cite{Kostelecky}, loop quantum gravity  \cite{Alfaro}, noncommutative geometry \cite{Carroll}, etc. Moreover, some Gedanken experiments in the spirit of black hole physics have also supported the idea of existence of a minimal measurable length \cite{Alfaro,Garay,Yan,Kostelecky,Gambini,Carroll,Burgess}. So, the existence of a minimal observable length is a common feature of all promising quantum-gravity candidates. In the quantum-gravity candidate, i.e. the Doubly Special Relativity (DSR), a minimal and observer-independent length is preserved at the expense of the Planck-scale violation  from the Lorentz invariance for dispersion relation \cite{Giovanni2}, which is named as the LIV-DR by us.
And, in another quantum-gravity candidate, i.e. the generalized uncertainty principle (GUP), a minimal observable length is preserved by modifying the uncertainty principle in quantum physics, where dispersion relation is not modified and the minimal length is described as an observer-dependent parameter due to the Lorentz symmetry \cite{Das,Pikovski,Pedram,Balasubramanian,Myung1,Ghosh,Gangopadhyay,
Hammad,Lidsey,Battisti,Nouicer,Miao,Adler,Nozari,Nozari1,Chen,Hossenfelder}. Obviously, in the LIV-DR and GUP candidates, quantum gravity are phenomenologically analyzed from different perspectives, so it is interesting to compare with their results. On the other hand, when the quantum-gravity effects are present, many physical phenomena that are absent at low energies appear at high energies, which maybe provide an effective window to solve some physical paradoxes that always occur in the semiclassical theory.
Motivated by these facts, our primary aim in this paper is to apply the LIV-DR with the dimensionless parameter $n=2$  and the ``sign of LIV'' $\eta_\pm = 1$, to phenomenologically study the effect of quantum gravity in the strong gravitational field. Specifically, we have studied the effect of the LIV-DR induced quantum gravity on the Schwarzschild black hole thermodynamics, and compared our results with the earlier findings by the GUP.

The outlines of this paper are listed as follows. In Sec. \ref{sec1}, we rewrite the Dirac equation with the inclusion of the LIV-DR in the curved spacetime, and study the LIV-DR effect on the emission rate of the Schwarzschild black hole in the tunneling framework\footnote{Here, the Hawking radiation is treated as a tunneling process at the horizon of black hole. This is a popular and intuitive method to study the Hawking radiation of black hole, also provides a possible solution for the black hole information loss after considering the black hole background as dynamical and the conservation of energy \cite{Hawking,Jiang,Parikh}.}. In Sec. \ref{sec2}, in the presence of the LIV-DR induced quantum gravity, we further analyze the thermodynamic properties at the final stage of the black hole evaporation, and reconsider the well-known singularity problem and critical problem that often occur in a semiclassical theory. Also, we compare our results with the GUP findings by Nozari \cite{Nozari} and Chen \cite{Chen}. Sec. \ref{sec4} ends up with conclusions and discussions.

\section{Quantum gravity and fermions' tunneling}
\label{sec1}
\setlength{\parindent}{2em}
In this section, we attempt to study the effect of the LIV-DR induced quantum gravity on fermions' tunneling radiation. Before that, we should first rewrite the Dirac equation with the inclusion of the LIV-DR. In the presence of the LIV-DR (\ref{Jeq1}), it is easy to write the modified Dirac equation for spin-1/2 particles as \footnote{We note that, this observation extends to the Dirac equation (\ref{Jeq2}), which arises from a very particular (D=5$\rightarrow$the dimension 5 operators) term in the SME, however many other choices are possible \cite{add1,add2}.} \cite{Kruglov1}
\begin{equation}
\left[ {{{\overline \gamma  }^\mu }{\partial _\mu } + \overline m  - i l_p \left( {{{\overline \gamma  }^t}{\partial _t}} \right)\left( {{{\overline \gamma  }^j}{\partial _j}} \right)} \right]\Psi  = 0. \label{Jeq2}
\end{equation}
Where, $\mu$ is the spacetime coordinates, $j$ is the space coordinates, and ${\overline \gamma  ^\mu }$ is the ordinary gamma matrix. It is easy to find, the Lorentz symmetry is broken by the additional term under boost transformations. This modified Dirac equation has been proved to be compatible with the quadratically suppressed dispersion relation (\ref{Jeq1}) when one adopts the wave function as the plane-wave solution $\Psi \left( x \right) = \Psi \left( p \right)\exp [{i}\left( {\overrightarrow p \cdot \overrightarrow x  - {p_0}\cdot{x_0}} \right)]$ to the equation (\ref{Jeq2}) \cite{Kruglov1}. Here, if we want to study fermions' tunneling radiation in the curved spacetime, we should first replace the gamma matrix and the partial derivative in (\ref{Jeq2}) with ${\overline \gamma  ^\mu } \to {\gamma ^\mu },{\partial _\mu } \to {D_\mu } = {\partial _\mu } + {\Omega _\mu } + \left( {{i \mathord{\left/{\vphantom {i \hbar }} \right. \kern-\nulldelimiterspace} \hbar }} \right)e{A_\mu }$, where ${\gamma ^\mu }$ satisfies the relation $\left\{ {{\gamma ^\mu },{\gamma ^\nu }} \right\} = {\gamma ^\mu }{\gamma ^\nu } + {\gamma ^\nu }{\gamma ^\mu } = 2{g^{\mu \nu }}I$, $e{A_\mu }$ is the charge term of the Dirac equation, ${\Omega _\mu }$ is the spin connection. So, in the curved spacetime, the deformed Dirac equation with the inclusion of the LIV-DR can be written as
\begin{equation}
\left[ {{\gamma ^\mu }{D_\mu } + \frac{m}{\hbar } - i \hbar l_p \left( {{\gamma ^t}{D_t}} \right)\left( {{\gamma ^j}{D_j}} \right)} \right]\Psi  = 0.  \label{Jeq3}
\end{equation}
Following the standard ansatz, it is necessary for us to rewrite the wave function of the Dirac equation as
\begin{equation}
\Psi  = \xi \left( {t,{x^j}} \right)\exp \left[ {\frac{i}{\hbar }S\left( {t,{x^j}} \right)} \right],  \label{Jeq4}
\end{equation}
where $S$ is the action of the tunneling fermions. Substituting this wave function into the deformed Dirac equation (\ref{Jeq3}) and carrying on the separation of variables as $S =  - \omega t + W( r ) + \Theta ( {\theta ,\varphi } )$ for the spherically symmetric space-time, we finally obtain the following motion for the action $S$, that is
\begin{equation}
\begin{aligned}
\left[i \gamma^{\mu}(\partial_{\mu}S+e A_{\mu})+ m - i l_p \gamma^t (\omega - e A_t) \gamma^j (\partial_{j}S + e A_j)\right] \xi (t,r,\theta,\varphi)=0,  \label{Jeq5}
\end{aligned}
\end{equation}
where ${\partial_t}S =-\omega$, and $\omega$ is the energy of the emitted Dirac particles, and the terms of $\hbar \Omega _\mu$ has been neglected at high energies. In this paper, we take the Schwarzschild black hole as an example to study the effect of the LIV-DR induced quantum gravity on fermions' tunneling radiation. For the Schwarzschild black hole, the line element of the spacetime can be written as $d{s^2} =  - f( r )d{t^2} + {{g( r )}^{-1}}d{r^2} + {r^2}\left( {d{\theta ^2} + {{\sin }^2}\theta d{\varphi ^2}} \right)$, where $f\left( r \right) = g\left( r \right) = 1 - {{2M}}/{r}$, and ${r_h} = 2M$ is the event horizon of the black hole. For the Dirac particles with spin $1/2$, there are two spin states, i.e. spin-up state ($\uparrow$) and spin-down state ($\downarrow$). In our case, we choose the spin-up state without loss of generality.  So, we have
\begin{equation}
\xi_{\uparrow} ( {t,r,\theta ,\varphi } ) = \left( {\begin{array}{*{20}{c}}
{A( {t,r,\theta ,\varphi } )\zeta_{\uparrow}}\\
{B( {t,r,\theta ,\varphi } )\zeta_{\uparrow}}\\
\end{array}} \right),\label{Jeq6}
\end{equation}
where $\zeta_{\uparrow}=\Big( {\begin{array}{*{20}{c}}
{1}\\
{0}\\
\end{array}} \Big)$ for the spin-up state. Now, we choose the suitable gamma matrixes to further simplify the Dirac equation (\ref{Jeq5}). For the Schwarzschild spacetime, we choose
\begin{equation}
\begin{array}{l}
{\gamma ^t} = \sqrt{ f^{-1}} \Big( {\begin{array}{*{20}{c}}
0&I\\
{ - I}&0
\end{array}} \Big),{\gamma ^r} = \sqrt g \Big( {\begin{array}{*{20}{c}}
0&{{\sigma ^3}}\\
{{\sigma ^3}}&0
\end{array}} \Big),\\
{\gamma ^\theta } = \sqrt {{g^{\theta \theta }}} \Big( {\begin{array}{*{20}{c}}
0&{{\sigma ^1}}\\
{{\sigma ^1}}&0
\end{array}} \Big),{\gamma ^\varphi } = \sqrt {{g^{\varphi \varphi }}} \Big( {\begin{array}{*{20}{c}}
0&{{\sigma ^2}}\\
{{\sigma ^2}}&0
\end{array}} \Big), \label{Jeq7}
\end{array}
\end{equation}
where, ${\sigma ^i}\left( {i = 1,2,3} \right)$ represent the Pauli matrixes. Substituting Eqs. (\ref{Jeq6}) and (\ref{Jeq7}) into Eq. (\ref{Jeq5}), we finally obtain four simplified equations with relation to functions $A$ and $B$. And, two of them are
\begin{equation}
B\Big( { - \frac{{i\omega }}{{\sqrt f }} + i\sqrt g {\partial _r}W} \Big) + A\Big( {m -  i l_p \omega {\partial _r}W} \Big) = 0,\label{Jeq8}
\end{equation}
\begin{equation}
A\Big( {\frac{{i\omega }}{{\sqrt f }} + i\sqrt g {\partial _r}W} \Big) + B\Big( {m +  i l_p \omega {\partial _r}W} \Big) = 0. \label{Jeq9}
\end{equation}
In Eqs. (\ref{Jeq8}) and (\ref{Jeq9}), if functions $A$ and $B$ are required to have non-trivial solutions, the determinant of the coefficient matrixes must be zero. That demands
\begin{equation}
{\partial _r}W( r ) =  \pm \sqrt {\Big( {\frac{{{\omega ^2}}}{{{f^2}}} - \frac{{{m^2}}}{f}} \Big)} \Big( 1-\frac{l_p^2 \omega^2}{2 f}\Big), \label{Jeq10}
\end{equation}
where, we have carried out the Taylor's expansion with respect to $l_p$, and the high-order terms, i.e. $O (l_p^4)$, have already been neglected.  When one performs the integral with respect to equation (\ref{Jeq10}), there is a pole ($r=r_h$) at the event horizon. To avoid it, we need to  take a suitable complex contour to bypass the pole. As described in \cite{res}, since the portion of the trajectory that starts outside the black hole and continues to the observer has no contribution to the final tunneling probability, the wave equation that contributes to the tunneling probability only come from the contour around the black hole horizon. In this case, by using the residue principle near the event horizon of the Schwarzschild black hole \cite{res}, we have
\begin{equation}
W_{ \pm } =  \pm 2i\pi \omega M\Big( {1 + l_p^2 \big( {\frac{{{m^2}}}{4} - {\omega ^2}} \big)} \Big ), \label{Jeq11}
\end{equation}
where the sign $\pm$ corresponds to the outgoing/ingoing Dirac particle across the event horizon of the black hole. Based on the WKB approximation, the relationship between the imaginary part of the action and the tunneling probability is given by $P = \exp \left( { - \frac{2}{\hbar }{\mathop{\rm Im}\nolimits} S} \right)$. Then, during the Dirac particle's tunneling across the black hole horizon, the total emission rate is
\begin{equation}
\begin{aligned}
\Gamma  &= \frac{{{P_{out}}}}{{{P_{in}}}} = \frac{{\exp ( { - 2{\mathop{\rm Im}\nolimits} {W_ + }} )}}{{\exp ( { - 2{\mathop{\rm Im}\nolimits} {W_ - }} )}} \\
&= \exp \Big[ { - 8\pi M\omega \Big( {1 - {l_p^2}\big( {{\omega ^2} - \frac{{{m^2}}}{4}} \big)} \Big)} \Big].\label{Jeq12}
\end{aligned}
\end{equation}
Clearly, a small correction to the tunneling rate has been exhibited by the effect of the LIV-DR induced quantum gravity. At first sight, this correction speeds up the black hole evaporation, which is consistent with the findings by Nozari \cite{Nozari,Nozari1} and Banerjee \cite{Rabin}, but is in contradiction with those by Chen \cite{Chen}. In \cite{Nozari,Nozari1,Rabin}, the effect of quantum gravity was present by demanding the Generalized Uncertainty Principle (GUP) with the incorporation of both a minimal observable length and a maximal momentum. Instead, the simplest GUP relation which implies only the appearance of a nonzero minimal length was incorporated in \cite{Chen}. Magueijo and Smolin have shown that in the context of the DSR, a test particle's momentum cannot be arbitrarily imprecise and therefore there is an upper bound for momentum fluctuations \cite{Smolin}. Then, it has been shown that this may lead to a maximal measurable momentum for a test particle \cite{Cortes}. That means that, in the GUP model, the appearance of the maximal momentum could provide a better description about the effect of quantum gravity. Combined with these facts, it also confirms from (\ref{Jeq12}) that, as a candidate for describing quantum gravity, the LIV-DR with the dimensionless parameter $n=2$  and the ``sign of LIV'' $\eta_\pm = 1$ can always do as well as the GUP with a minimal observable length and a maximal momentum. On the other hand, this correction (\ref{Jeq12}) due to the presence of the LIV-DR induced quantum gravity becomes drastic at the Planck scale, and the final stage of the black hole evaporation would exhibit some interesting properties, which will be detailed in Sec. \ref{sec2}. In addition, the quantum-gravity correction of the tunneling rate (\ref{Jeq12}) is related not only to the mass of black hole, but also to the Planck length and the emitted fermions' mass and energy. So, we see that in the presence of the quantum-gravity effect, the emission spectrum cannot be strictly thermal. This happens to coincide with the Parikh-Wilczek's observation \cite{Parikh}, where a leading correction to the tunneling rate has been present by the incorporation of the emitted particle's self-gravitational interaction. However, because of the lack of correlation between different emitted modes in the black hole emission spectrum, the form of the Parikh-Wilczek's correction is not adequate by itself to recover information \cite{Jr16}. This semiclassical correction with the incorporation of a minimal measurable length and possible resolution of the information loss problem in this framework has been studied in \cite{Nozari1}. Later, in the tunneling framework, further research with the GUP including all natural cutoffs shows that information emerges continuously during the evaporation process at the quantum-gravity level \cite{Nozari}. This property has the potential to answer some questions about the black hole information loss problem and provides a more realistic background for treating the black hole evaporation in its final stage. In fact, these observations only provide evidence of correlations between the two tunneling particles in the presence of quantum gravity, but they are not adequate by themselves to recover information. Sec. \ref{sec4} ends up with some discussions on the information loss.

\section{Remnant values of temperature, mass and entropy}
\label{sec2}
\begin{figure}[H]
  \centering
  \includegraphics[width=0.6\textwidth]{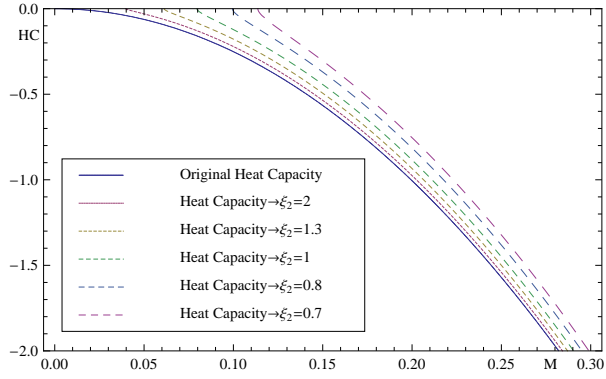}\\
  \caption{\scriptsize{: The LIV-DR induced black hole heat capacity versus its mass.}}\label{figure1}
\end{figure}

\begin{figure}[t]
  \centering
  \includegraphics[width=0.6\textwidth]{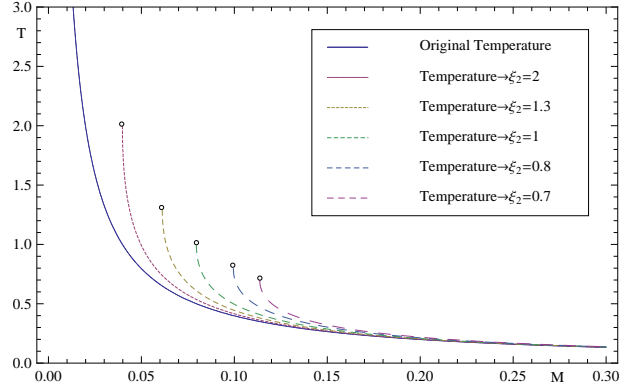}\\
  \caption{\scriptsize{: The LIV-DR induced black hole temperature versus its mass.}}\label{fig2}
\end{figure}

\begin{figure}[H]
  \centering
  \includegraphics[width=0.6\textwidth]{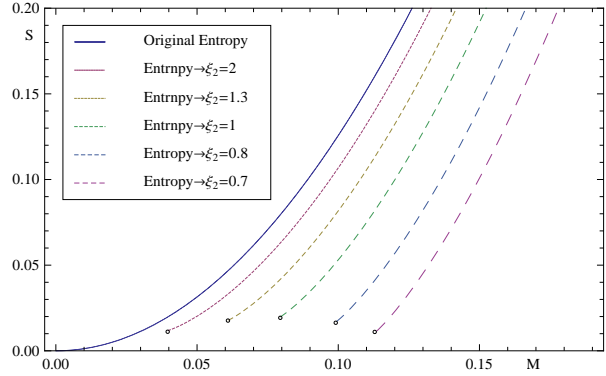}\\
  \caption{\scriptsize{: The LIV-DR induced black hole entropy versus its mass.}}\label{fig3}
\end{figure}

\begin{figure}[H]
  \centering
  \includegraphics[width=0.6\textwidth]{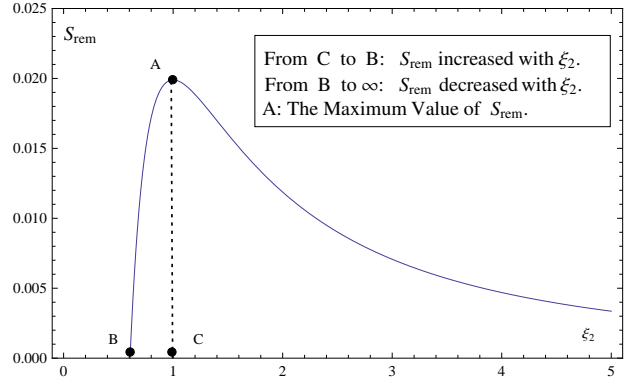}\\
  \caption{\scriptsize{: The LIV-DR induced remnant entropy versus $\xi_2$. }}\label{fig7}
\end{figure}

\setlength{\parindent}{2em}
At the quantum-gravity level, an interesting attempt is to observe the Planck-scale physics. At the final stage of the black hole evaporation, some exciting findings due to the presence of the quantum-gravity effect would be exhibited. In this section, with the inclusion of the LIV-DR induced quantum-gravity effect, we attempt to obtain the Planck-scale thermodynamic quantities at the final stage of the black hole evaporation. Before that, we should first find the quantum-gravity induced thermodynamic relations. As defined by \cite{Chen,Vagenas,Jr13}, using the principle of ``detailed balance'' for the emission rate (\ref{Jeq12}) yields the effective temperature of the Schwarzschild black hole given by
\begin{equation}
T = {T_h}\left( {1 + {l_p^2}\Big( {{\omega ^2} - \frac{{{m^2}}}{4}} \Big)} \right), \label{Jeq13}
\end{equation}
where $l_p={L_p}/{\xi_2}$ and $ {L_p} = {1}/{M_p}$ is the Planck length, $T_h = {M_p^2}/({8 \pi M k_B})$ is the standard semiclassical temperature of the black hole and other terms are the corrections due to the quantum-gravity effect, and we use the relations $c=\hbar=1$ and $\hbar=L_p M_p=G M_p^2$. To further facilitate our calculations, deforming the equation (\ref{Jeq13}) at the final stage of black hole evaporation, we have
\begin{equation}
{M} = \frac{{M_p^2}}{{8\pi T k_B}}\left( {1 + {\frac{k_B^2{T^2} L_p^2}{\xi_2^2}}} \right), \label{Jeq15}
\end{equation}
where, we have approximately replaced $\omega $ in (\ref{Jeq13}) with the characteristic energy of the emitted particle \cite{Rabin,Adler}. For the particle with temperature $T$, its characteristic energy is given by $ k_BT $ \cite{Rabin}.
Normally, a black hole has a negative heat capacity, so its temperature would increase when the black hole loses mass and energy by the emission process. In Fig. \ref{figure1}, we have shown the LIV-DR induced black hole heat capacity versus its mass in the different values for $\xi_2$. Obviously, the heat capacity of the black hole under the quantum-gravity corrections is bigger than that of the semiclassical case. And with the decrease of the parameter $\xi_2$, the corrected heat capacity is much closer to the semiclassical one.
When considering the quantum-gravity effect into the emission process, the negative heat capacity increases monotonically as the energy gradually reaches the Planck energy. There is a point at which the heat capacity vanishes. The corresponding temperature is considered to be the maximum temperature attainable by the black hole evaporation. So, at the final stage of the black hole evaporation, there is no further change of black hole mass with its temperature. This means that the heat capacity of the black hole defined by ${C} = \frac{{dM}}{{dT}}$ becomes zero, at which the emission process ends with a finite remnant mass and temperature.  According to Eq. (\ref{Jeq15}), the final stage of the black hole evaporation leaves the \textbf{\textsl{remnant temperature}} given by
\begin{equation}
{{{T}}_{rem}} = \frac{{\xi _2 {M_p}}}{{{k_B}}},  \label{Jeq16}
\end{equation}
and the corresponding \textbf{\textsl{remnant mass}} is given by
\begin{equation}
{{{M}}_{rem}} = \frac{{{M_p}}}{{4\pi \xi_2}}. \label{Jeq17}
\end{equation}
Alternatively, the remnant mass can also be obtained by minimising the black hole entropy, that is $\frac{{dS}}{{dM}}=0$, and the second derivative $\frac{d^2S}{dM^2}>0$. In Fig. \ref{fig2}, we have shown the LIV-DR induced black hole temperature versus its mass in the different values for $\xi_2$. Obviously, in the presence of the LIV-DR effect, the quantum-gravity corrected temperature is higher than the semiclassical case. And, with the increase of the parameter $\xi_2$, the corrected temperature is much closer to the semiclassical temperature, meanwhile the final stage of the black hole evaporation leaves a higher remnant temperature and a smaller remnant mass.
Next, we attempt to obtain the modified black hole entropy in the presence of the LIV-DR effect. According to the first law of the black hole thermodynamics, the black hole entropy is given by
\begin{eqnarray}
{{S}} &=& \int \frac{dM}{T} = \int \frac{CdT}{T}\nonumber\\
&=& \frac{k_B}{16\pi}\Big[\Big(\frac{M_p}{k_BT}\Big)^2 +\frac{1}{\xi_2^{2}} ln\Big(\frac{k_BT}{M_p}\Big)^2\Big].\label{Jeq18}
\end{eqnarray}
So, at the final stage of the black hole evaporation, the \textbf{\textsl{remnant entropy}} is left by
\begin{equation}
{{{S}}_{rem}} = \frac{k_B}{16\pi \xi_2^2}\Big(1- ln \frac{1}{\xi_2^2} \Big). \label{JJeq17}
\end{equation}
In (\ref{Jeq18}), the corrected black hole entropy is expressed in terms of the corrected temperature, so it is inconvenient to observe the quantum-gravity corrections to the semiclassical Bekenstein-Hawking entropy. To specifically exhibit the quantum-gravity induced black hole entropy in terms of the semiclassical Bekenstein-Hawking entropy, we should first find an expression for $T^2$ in terms of $M$. We can do this by squaring (\ref{Jeq15}), that is
\begin{eqnarray}
\Big(\frac{k_B T}{M_p}\Big)^2 = \frac{\Big[\Big(\frac{8\pi M}{M_p}\Big)^2-\frac{2}{\xi_2^{2}}\Big]\pm \sqrt{\Big[\Big(\frac{8\pi M}{M_p}\Big)^2-\frac{2}{\xi_2^{2}}\Big]^2-\frac{4}{\xi_2^{4}}}}{2\xi_2^{-4}}. \label{Jeq19}
\end{eqnarray}
Here, only the part with the ($-$) sign is acceptable, since the ($+$) part can not produce the semiclassical Mass-Temperature result in the absence of the quantum-gravity effect (i.e. $\xi_2^{-2}=0$). Now, rearranging (\ref{Jeq19}) with binomial expansion, we have
\begin{eqnarray}
\Big(\frac{k_B T}{M_p}\Big)^2 = \frac{1}{\Big[\Big(\frac{8\pi M}{M_p}\Big)^2-\frac{2}{\xi_2^{2}}\Big]}\left[1+\frac{\xi_2^{-4}}{\Big[\Big(\frac{8\pi M}{M_p}\Big)^2-\frac{2}{\xi_2^{2}}\Big]^2}+\cdot\cdot\cdot\right]. \label{Jeq20}
\end{eqnarray}
Substituting (\ref{Jeq20}) into (\ref{Jeq18}), we can obtain the modified black hole entropy
\begin{eqnarray}
&&\frac{S}{k_B} =\Big(\frac{S_{BH}}{k_B}-\frac{2}{16\pi \xi_2^{2}}\Big)-\frac{1}{16\pi \xi_2^{2}}ln \Big(\frac{S_{BH}}{k_B}-\frac{2}{16\pi \xi_2^{2}}\Big)\nonumber\\
&&+\sum_{j=0}^\infty c_j(\xi_2^{-2})\Big(\frac{S_{BH}}{k_B}-\frac{2}{16\pi \xi_2^{2}}\Big)^{-j}-\frac{1}{16\pi \xi_2^{2}} ln 16\pi, \label{Jeq21}
 \end{eqnarray}
where the semiclassical Bekenstein-Hawking entropy is given by $S_{BH}=k_B\frac{4\pi M^2}{M_p^2}$, and the coefficients $c_j$ are the functions about $\xi_2^{-2}$. This is the modified black hole entropy in the presence of the LIV-DR induced quantum-gravity effect, and the semiclassical Bekenstein-Hawking entropy would be reproduced when the quantum-gravity effect is absent (i.e. $\xi_2^{-2}=0$). We subsequently attempt to obtain the area theorem from the expression of the modified black hole entropy (\ref{Jeq21}). Before that, we introduce a new variable $\widetilde{\textbf{A}}$ defined by $\widetilde{\textbf{A}}=16\pi G^2 M^2-\frac{2}{4\pi \xi_2^{2} }G^2M_p^2=\textbf{A}-\frac{2}{4\pi \xi_2^{2} } L_p^2$, which is named as
the \textbf{\textsl{ reduced area}}, and $\textbf{A}=16\pi G^2 M^2$ is the usual area of the black hole horizon. In terms of the reduced area, the modified black hole entropy can be written in a familiar form
\begin{eqnarray}
\frac{S}{k_B}=\frac{\widetilde{\textbf{A}}}{4L_p^2}-\frac{1}{16\pi \xi_2^{2}}ln \Big(\frac{\widetilde{\textbf{A}}}{4L_p^2}\Big) + \sum_{j=0}^\infty c_j(\xi_2^{-2})\Big(\frac{\widetilde{\textbf{A}}}{4L_p^2}\Big)^{-j}-\frac{1}{16\pi \xi_2^{2}} ln 16\pi. \label{Jeq22}
 \end{eqnarray}
This is the area theorem in the presence of the LIV-DR induced quantum-gravity effect. The usual Bekenstein-Hawking semiclassical area law is reproduced for $\xi_2^{-2}=0$. This quantum-gravity induced area theorem looks like the standard modified area theorem \cite{Kaul,Vagenas,Jr1,Jr2,Jr4,Jr5,Jr7,Jr8,Jr9,Jr10,Jr11,Jr12}, with the role of the usual area $\textbf{A}$ being played by the reduced area $\widetilde{\textbf{A}}$. We also note that as usual the leading order correction term to the black hole entropy has a logarithmic nature, which is consistent with the earlier findings by the Loop Quantum Gravity considerations \cite{Vagenas,Jr1,Jr12}, the field theory calculations \cite{Jr4,Jr5}, the quantum geometry method \cite{Kaul,Jr11}, and the statistical method \cite{Jr8,Jr9}. The higher order corrections involve inverse powers of the reduced area $\widetilde{\textbf{A}}$. In Fig. \ref{fig3}, we have shown the LIV-DR induced black hole entropy versus its mass in the different values for $\xi_2$. Obviously, the quantum-gravity corrected entropy is lower than the semiclassical case. And, with the increase of the parameter $\xi_2$, the corrected entropy is much closer to the semiclassical entropy.
Also, in Fig. \ref{fig7}, we have shown the LIV-DR induced remnant entropy versus $\xi_2$. Obviously, when $0.6065\leq \xi_2 \leq 1 $, the remnant entropy increases with the increasing parameter $\xi_2$. And, when $\xi_2>1$, the remnant entropy decreases with the increasing parameter $\xi_2$. At the point $\xi_2=1$, the remnant entropy reaches maximum.
When the reduced area is zero (i.e. $\widetilde{\textbf{A}}=0$), there is a singularity for the modified black hole entropy (\ref{Jeq22}), which corresponds to the \textbf{\textsl{singular mass}} given by
\begin{equation}
M_{sin}=\frac{\sqrt{2}{{M_p}}}{{8\pi \xi_2}}.\label{Jeq23}
 \end{equation}
On the other hand, the \textsl{\textbf{critical mass}} below which the temperature (\ref{Jeq19}) becomes a complex quantity is given by
\begin{equation}
M_{cri}=\frac{{{M_p}}}{{4\pi \xi_2}}.\label{Jeq24}
 \end{equation}

In a word, the LIV-DR induced quantum-gravity effect speeds up the black hole evaporation, and its corresponding black hole entropy undergoes a leading logarithmic correction to the ``reduced Bekenstein-Hawking entropy'', which was found earlier in \cite{Kaul,Vagenas,Jr1,Jr2,Jr4,Jr5,Jr7,Jr8,Jr9,Jr10,Jr11,Jr12} with the role of the reduced area $\widetilde{\textbf{A}}$ being played by the usual area $\textbf{A}$. At the final stage of the black hole evaporation when the heat capacity reaches zero, the quantum-gravity effect stops the further collapse of the black hole with the remnant mass $M_{rem}$, the remnant temperature $T_{rem}$ and the remnant entropy $S_{rem}$. We also note that the remnant mass $M_{rem}$ is greater than the singular mass $M_{sin}$, but equal to the critical mass $M_{cri}$. So, during the black hole evaporation, we can easily find that the singularity problem is naturally avoided, and the reduced area is always be positive. Meanwhile, we manage to avoid the critical problem of dealing with the complex values for the thermodynamic entities. Consequently, the ill defined situations are naturally bypassed in the presence of the LIV-DR induced quantum-gravity effect. In addition, we have shown in Fig. \ref{figure1}, Fig. \ref{fig2} and Fig. \ref{fig3} that
the quantum-gravity effect lowers the black hole entropy, but elevates the black hole heat capacity and temperature. And, with the increase of the parameter $\xi_2$, the quantum-gravity corrected heat capacity, temperature and entropy are all much closer to the semiclassical counterparts, meanwhile the final stage of the black hole evaporation leaves a smaller remnant mass and a higher remnant temperature. In Fig. \ref{fig7}, we have shown the remnant entropy increases with the increasing parameter $0.6065\leq \xi_2 \leq 1$, and decreases with the increasing parameter $\xi_2>1$. At the point $\xi_2=1$, the remnant entropy reaches maximum.

\begin{figure}[H]
  \centering
  \includegraphics[width=0.6\textwidth]{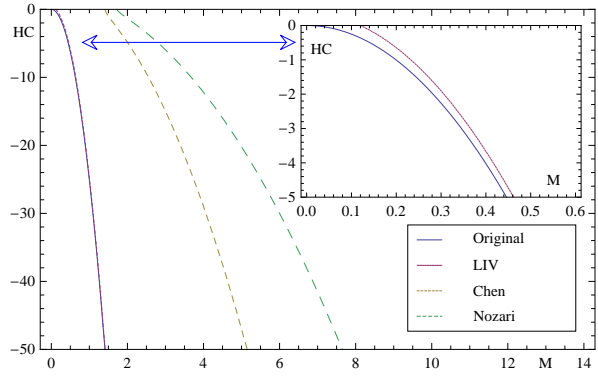}\\
  \caption{\scriptsize{: The black hole's modified heat capacity versus its mass when employing the GUP parameters $\beta_0=\alpha=1$ and the LIV parameter $\xi_2 =1$.}}\label{fig4}
\end{figure}

\begin{figure}[H]
  \centering
  \includegraphics[width=0.65\textwidth]{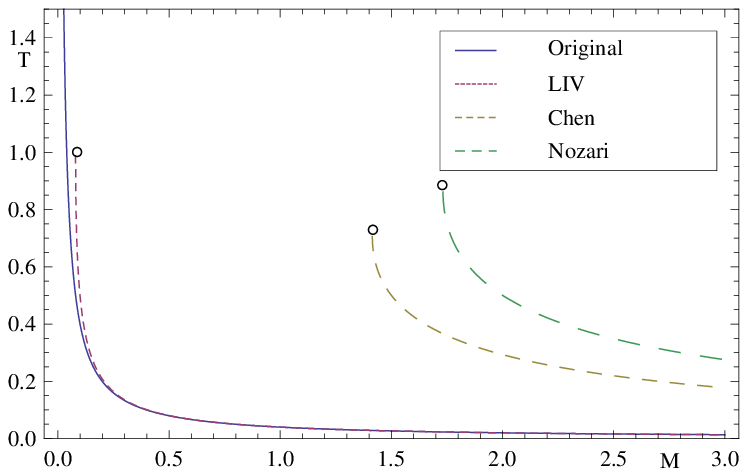}\\
  \caption{\scriptsize{:  The black hole's modified temperature versus its mass when employing the GUP parameters $\beta_0=\alpha=1$ and the LIV parameter $\xi_2 =1$.
  }}\label{fig5}
\end{figure}

\begin{figure}[H]
  \centering
  \includegraphics[width=0.65\textwidth]{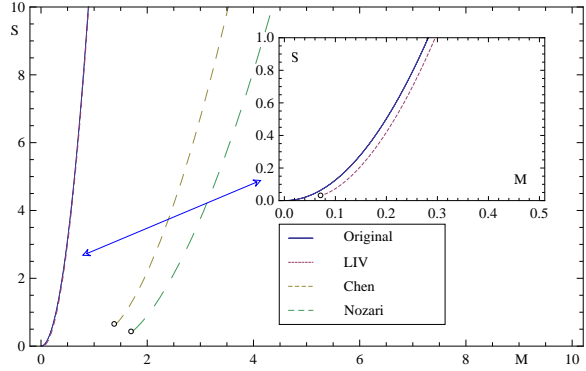}\\
  \caption{\scriptsize{: The black hole's modified entropy versus its mass when employing the GUP parameters $\beta_0=\alpha=1$ and the LIV parameter $\xi_2 =1$.}}\label{fig6}
\end{figure}

To put our results in a proper perspective, let us compare with the earlier findings by another quantum gravity candidates, i.e. the generalized uncertainty principle (GUP) \cite{Das,Pikovski,Pedram,Balasubramanian,Myung1,Ghosh,Gangopadhyay,
Hammad,Lidsey,Battisti,Nouicer,Miao,Adler,Nozari,Nozari1,Chen,Hossenfelder,Jr15}, where the dispersion relation is not modified and the minimal length is instead described as an observer-dependent parameter due to the Lorentz symmetry. In Fig. \ref{fig4}, Fig. \ref{fig5} and Fig. \ref{fig6}\footnote{In Figs.\ref{fig4},\ref{fig5},\ref{fig6}:\\
(\textbf{a}), to plot the curve ``Chen'', we have applied the mass-temperature relationship $ 2 \beta_0 T^2 - 2 \beta_0 M T +1 =0 $, which comes from the equation $( M - dM )( 1 + \frac{2 \beta_0 \omega^2}{M_p^2} ) \simeq M $ under the situation $dM = \omega = k_B T$ in Refs. \cite{Chen};\\
(\textbf{b}), to plot the curve ``Nozari'', we have used the mass-temperature relationship with the form of $ \frac{4}{3} \alpha^2 T^2 - \frac{4}{3} \alpha^2 M T +1 =0$, since the equation $( M - dM )( 1 + \frac{4 \alpha^2 \omega^2}{3 M_p^2} ) \simeq M$ was obtained from the tunneling rate $ \Gamma \sim exp \big[ -8 \pi M \omega + 4 \pi \omega^2 -2\pi \alpha L_p^2 \omega^3 (\frac{16}{3} - 5 \omega ) + O(\alpha^2 L_p^4 ) \big]$ in Refs. \cite{Nozari}, in which the background variation in black hole evaporation is neglected and only the second-order correction of $\alpha$ is considered. }, when employing the GUP parameters $\beta_0=\alpha=1$ and the LIV parameter $\xi_2 =1$, we have specifically compared the LIV-DR induced results with the GUP induced findings by Nozari \cite{Nozari,Nozari1} and Chen \cite{Chen} without loss of generality. In Fig. \ref{fig4}, we have found that the LIV-DR induced heat capacity is higher than the semiclassical one, but is lower than the GUP induced one. In Fig. \ref{fig5}, we have shown that the LIV-DR induced black hole temperature is higher than the semiclassical one, but is lower than that of the GUP case. Meanwhile, at the final stage of the black hole evaporation, the LIV-DR induced quantum gravity effect leaves a smaller remnant mass, and a higher remnant temperature than the case of the GUP induced quantum gravity. In Fig. \ref{fig6}, we have noted that the LIV-DR induced black hole entropy is lower than the semiclassical one, but is higher than that of the GUP case. And, at the final stage of the black hole evaporation, the remnant entropy induced by the LIV-DR quantum gravity is smaller than that induced by the GUP quantum gravity. Here, it is noted that our treatment for the GUP \cite{Nozari,Nozari1,Chen} is universal since it contains all natural cutoffs as a minimal length, a minimal momentum and a maximal momentum.


\section{Conclusions and Discussions}\label{sec4}

\setlength{\parindent}{2em}
In this paper, we have applied the Lorentz-Invariance-Violation (LIV) class of dispersion relations (DR) with the dimensionless parameter $n=2$  and the ``sign of LIV'' $\eta_\pm = 1$, to phenomenologically study the effect of quantum gravity in the strong gravitational field. Specifically, we have studied the effect of the LIV-DR induced quantum gravity on the Schwarzschild black hole thermodynamics. First, we have written out the modified Dirac equation in accordance with the deformed dispersion relation with the Lorentz invariance violation.
Then, in the tunneling framework, we have applied the deformed Dirac equation to study the effect of quantum gravity on the black hole emission. Finally, we have shown some peculiar properties of the LIV-DR induced quantum gravity at the final stage of the black hole evaporation, and compared them with the GUP induced observations. The result shows that, the effect of the LIV-DR induced quantum gravity speeds up the black hole evaporation, and its corresponding black hole entropy undergoes a leading logarithmic correction to the ``reduced Bekenstein-Hawking entropy'', and the ill defined situations (i.e. the singularity problem and the critical problem) are naturally bypassed in the presence of this quantum-gravity effect.
Our work once again provides a piece of evidence for the dimensionless parameter $n=2$ in the LIV-DR. Also, the result shows that, at the same quantum-gravity dependent parameters, the LIV-DR induced black hole heat capacity and temperature are lower than the GUP induced ones, but the LIV-DR induced black hole entropy is higher than the GUP induced one, meanwhile at the final stage of the black hole evaporation, the LIV-DR effect leaves a smaller remnant mass, a higher remnant temperature, and a smaller remnant entropy than the case of the GUP.

In the standard view of black hole thermodynamics, a black hole should emit blackbody radiation, thereby becoming lighter and hotter, and so on, leading to an explosive end when the mass approaches zero. However, when including of the effect of quantum gravity during the emission process, a black hole stops a further collapse at a remnant mass, temperature and entropy, and becomes an inert remnant, possessing only gravitational interactions. Obviously, the black hole's remnants contain information of quantum gravity, and need not have a classical black hole horizon structure. In this paper, we have shown that, when the quantum-gravity dependent parameters $\xi_2^{-1}=\beta_0=\alpha$, the LIV-DR induced remnant entropy is smaller than the GUP induced one, which suggests that, the final structure of the black hole is unstable in the GUP induced remnant entropy, and can further evolve till its final entropy at least reaches the LIV-DR induced remnant entropy. This also suggests that in the information theory, if the remnant entropy (information) is used to describe the basic structure of quantum gravity (In fact, this is well done in accordance with the following discussion on the information loss), the LIV-DR induced remnants contain more information of quantum gravity than the GUP induced one. So, by analysing the influences of the two quantum-gravity candidates on the final stage of black hole evaporation, we find that, the LIV-DR acts as a more suitable candidate for describing quantum gravity in comparation with the GUP. However, it is noteworthy that, the above discussions are only validated when the LIV-DR parameter is exactly equal to the GUP parameter. Normally, when $\xi_2^{-1}> \beta_0, \alpha$, we can always adjust the model-dependent parameters to make the LIV-DR and GUP induced remnant entropy equal. We also note that, when the GUP includes all natural cutoffs and higher order corrections \cite{Rabin}, its induced remnant entropy is exactly equal to the LIV-DR induced one (\ref{JJeq17}) when $\xi_2^{-1}=a_1'$. This shows that at the final stage of black hole evaporation, the GUP with all natural cutoffs and higher order corrections contains the same amount of information of quantum gravity as the LIV-DR. On the other hand, the third law of thermodynamics demands that the remnant entropy at the final stage of black hole evaporation should be greater than zero (i.e. $S_{rem}\geq 0$), which, in our case, determines the LIV-DR parameter $\xi_2\geq 0.6065$. Obviously, this parameter range is within a range of parameters from flaring active galactic nucleus(AGNs) for $n = 2$ case ${\xi _2} \ge {10^{ - 9}}$  and gamma-ray bursts(GRBs) for $n = 1$ case ${\xi _1} \ge 0.01$.

The information loss paradox during the Hawking radiation is an outstanding issue for the black hole physics. As a heuristic progress, the semiclassical method that treats the Hawking radiation as tunneling, has been proposed by Parikh and Wilczek to recover the unitarity for the black hole emission \cite{Parikh}. However, further research also by Parikh has shown that there are lack of correlations between the tunneling radiation of two particles \cite{Jr16}. This means that, although the tunneling picture might recover the unitarity for the black hole emission, it is not sufficient by itself to relay information. Later on, with the inclusion of quantum gravity, some attempts have been proposed to recover information in the tunneling picture \cite{Nozari,Nozari1}. In fact, these observations only provide evidence of correlations between the two tunneling particles in the presence of quantum gravity, but they are not adequate by themselves to recover information. Recently, an interesting observation has shown that the Hawking radiation as tunneling is an entropy conservation process, which leads naturally to the conclusion that the process of Hawking radiation is unitary, and no information loss occurs \cite{Zhang1}. When the effect of quantum gravity is included, more information would be present in the black hole radiation, but we assert that the Hawking radiation as tunneling is still an entropy conservation process, which will be reported by us in the near future. So, at the final stage of black hole evaporation, the inert remnant contains information of quantum gravity, possessing only gravitational interactions.

\textbf{Acknowledgments}

This work is supported by the Program for NCET-12-1060, by the Sichuan Youth Science and Technology Foundation with Grant No. 2011JQ0019, and by FANEDD with Grant No. 201319, and by the Innovative Research Team in College of Sichuan Province with Grant No. 13TD0003, and by Sichuan Natural Science Foundation with Grant No. 16ZB0178, and by the starting fund of China West Normal University with Grant No.14D014.


\end{document}